\documentclass[aps,pra,twocolumn,superscriptaddress,showpacs,floatfix]{revtex4}
\usepackage{graphicx}
\usepackage{color}
\usepackage{bm}
\usepackage{amsmath}

\newcommand{\fig}[1]{Fig.~(\ref{#1})}
\newcommand{\be}[1]{\begin{equation}\label{#1}}
\newcommand{\ee}{\end{equation}}

\renewcommand{\vec}[1]{{\boldsymbol #1}}

\begin{document}

\title{Intensity dependence of strong field double ionization mechanisms: from field-assisted recollision ionization to recollision-assisted field ionization}
%\title{Intensity dependence of ionization mechanisms for infrared frequencies in the strong field double ionization of diatomic molecules}

\author{A. Emmanouilidou}
\affiliation{Department of Physics and Astronomy, University College London\\
Gower Street, London WC1E 6BT, United Kingdom\\
and Chemistry Department, University of Massachusetts at Amherst, Amherst, Massachusetts, 01003} \date{\today}
\author{A. Staudte}
\affiliation{Steacie Institute for Molecular Sciences, National Research Council of Canada, 100 Sussex Drive, Ottawa, Ontario K1A 0R6, Canada}

\date{\today}

\begin{abstract}
Using a three-dimensional quasiclassical technique we explore
molecular double ionization by a linearly polarized, infrared
(800~nm) and ultrashort (6~fs) laser pulse. We first focus on
intensities corresponding to the tunneling regime and identify the
main ionization mechanisms in this regime. We devise a selection of
observables, such as, the correlated momenta and the sum of the
momenta parallel to the laser field as a function of the
inter-electronic angle of escape where all the main mechanisms have
distinct traces. Secondly, we address intensities above but close to
the over-the-barrier intensity regime. We find a surprising
anti-correlation of electron momenta similar to the experimental
observations reported in Phys. Rev. Lett. \textbf{101}, 053001
(2008). There, however, the anti-correlation was observed in very
low intensities corresponding to the multiphoton regime. We discuss
the mechanism responsible for the anti-parallel two-electron escape.

\end{abstract}


 \maketitle

\section{Introduction}

In the past two decades electron correlation has been established as
the underlying mechanism for many important phenomena arising from
the interaction of strong laser pulses with matter. One of these
phenomena is the dramatically enhanced multiple ionization yield of
atoms (e.g. \cite{WalkerPRL94}) and molecules (e.g.
\cite{AlnaserPRL2003}) in intense laser pulses. Also called
non-sequential multiple ionization, it is the laser driven
recollision \cite{CorkumPRL93,SchaferPRL93} of a field ionized
electron with its parent ion that governs the multiple ionization
process up to a certain threshold intensity \cite{RudenkoJPB2008}.
Beyond the threshold intensity the contribution of recollision to
multiple ionization vanishes and is replaced by a series of
independent, sequential field ionizations. Using coincidence imaging
techniques such as COLTRIMS many experiments have succeeded in
obtaining highly differential kinematical details of electron
correlation in the non-sequential intensity regime (e.g.
\cite{WeberNature2000,EreminaJPB2003,JesusJESRP2004,WeckenbrockPRL2004,StaudtePRL2007b,RudenkoPRL2007,LiuPRL2008}).
However, for higher intensities the available experimental data is
much less abundant. Correspondingly, the majority of theoretical
work has concentrated on the non-sequential intensity regime (e.g.
\cite{HoPRA2005,HaanPRL2006,BondarPRA2009}).

\begin{figure}
  \includegraphics[width=0.35\textwidth,keepaspectratio]{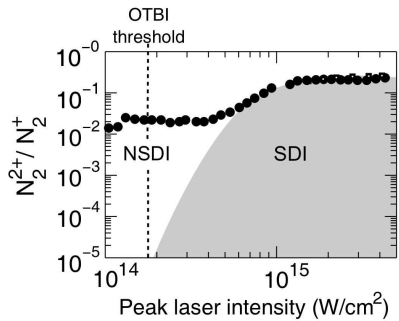}
    \caption{\label{fig1:CornaggiaN2} Ratio of N$_2$ double and single ionization yield
    as a function of laser intensity. Data points were experimentally determined in \cite{CornaggiaPRA2000}.
    At lower intensities non-sequential double ionization (NSDI) is the governing DI mechanism, though at higher
    intensities sequential double ionization (SDI) eventually dominates. The threshold for over-the-barrier
    ionization (OTBI) of N$_2$ is about $2\times10^{14}\mathrm{W/cm}^2$.}
\end{figure}

Here, we report a classical study of electron correlation in N$_2$
molecules subjected to strong laser pulses at intensities well
within the non-sequential double ionization regime (as indicated in
Fig.~\ref{fig1:CornaggiaN2}), as well as for intensities corresponding to
the transition from pure tunneling to over-the-barrier ionization.

First, we consider the intensity regime below the classical
over-the-barrier intensity. That is, one electron tunnels in the
field-lowered Coulomb potential, then accelerates in the laser field
and finally returns to the core to transfer energy to the remaining
electron. In this intensity regime we find that the time of minimum
electron approach is close to a zero of the field in agreement with
the re-collision model \cite{CorkumPRL93}. We differentiate the main
energy transfer mechanisms in the re-collision of the free electron
with the parent ion using the time delay between re-collision and
ionization of the second electron. Furthermore, we devise a set of
asymptotic observables where these mechanisms have distinct traces,
such as the sum of the momenta parallel to the laser field as a
function of the inter-electronic angle of escape.

Second, we consider intensities above, but close to, the
over-the-barrier intensity threshold. This is an intermediate regime
where a transition takes place from correlated, or non-sequential,
to independent, or sequential double ionization. We find that the
instant of re-collision shifts to a maximum of the field in contrast
to the common picture of non-sequential double ionization (NSDI)
where the time of re-collision is close to a zero of the field.
Furthermore, we find that the two electrons asymptotically depart in
opposite directions. This is quite unexpected since for very high
laser intensities one expects that both electrons ionize almost
immediately parallel to the laser field and thus parallel to each
other.  We discuss how the escape of both electrons in opposite
directions can be partly attributed to the laser field significantly
interacting with both electrons before the re-collision --- in
contrast to smaller intensities where the effect of the field on the
remaining electron can be neglected; another important factor is
that the remaining electron is still bound and thus the effect of
the Coulomb potential can not be neglected.

Addressing the double ionization of strongly driven systems
with first-principle quantum mechanical calculations in all three spatial
dimensions (3-d) is an immense task. Currently, 3-d ab-initio
quantum mechanical calculations are only available for the driven He
atom \cite{ParkerPRL2006}. To cope with the highly complex task of
tackling the double ionization of diatomic molecules many studies
use numerical quantum approaches of reduced dimensionality (e.g.
\cite{BandraukPRA2005}). Others use judiciously chosen quantum
models of reduced dimensionality (e.g. \cite{BaierPRA2008}), while
some use semi-analytical quantum approaches in the framework of the
so-called Strong-Field Approximation (SFA) (e.g.
\cite{FigueiraPRA2008}), not fully accounting for the Coulomb
singularity.

We use a 3-d quasiclassical technique that we developed for
conservative systems, i.e. the single photon multiple ionization of
atomic systems \cite{EmmanouilidouJPB2006,EmmanouilidouPRA2007,EmmanouilidouPRA2007b}.
Recently, we extended this technique to non-conservative systems to
treat the correlated electron dynamics of the driven He atom
\cite{EmmanouilidouPRA2008}. Here, we further build on this work by
tackling more than one atomic centers. Our method is numerically
very efficient and treats the Coulomb singularity with no
approximation in contrast to techniques that use ``soft-core"
potentials. Our method is also important for describing accurately
effects such as the striking so-called ``finger-like" structure
which was recently observed in He
\cite{StaudtePRL2007b,RudenkoPRL2007,EmmanouilidouPRA2008,YePRL2008}.
This structure was attributed to the strong interaction of the
rescattering electron with the core---backscattering. In addition,
accounting for the Coulomb singularity will be very important in
pump-probe set-ups where VUV or XUV pulses are used. For these high
frequency pulses, the excursion parameter of the electronic motion
is smaller than the atomic dimensions making it very important to
accurately incorporate the Coulomb potential.

\section{Method}

Our 3-d quasiclassical model entails the following steps: We first
set-up the initial phase space distribution of the two ``active"
electrons in the N$_{2}$ diatomic molecule. Here, we consider only
parallel alignment between the molecular axis and the laser electric
field. At intensities in the tunneling regime we assume that one
electron tunnels through the field-lowered Coulomb potential. For
the tunneling rate one can use quantum mechanical or semiclassical
formulas for diatomic molecules (see, e.g.
\cite{TongPRA2002,LitvinyukPRL2003,KjeldsenJPB2004, LiPRA2007}). We
use the rate provided in ref. \cite{LiPRA2007}. The longitudinal
momentum is zero while the transverse one is provided by a Gaussian
distribution \cite{LiuPRL2007}. This description is valid as long as
the potential barrier is not completely suppressed by the
instantaneous laser field $E(t)=E_0(t) \cos (\omega t )$. We
consider the usual laser wavelength of 800~nm, corresponding to
$\omega=0.057\mathrm{a.u.}$ (a.u. - atomic units). In our
simulation the pulse envelope $E_{0}(t)$ is defined as
$E_0(t)=E_{0}$ for $0<t<T$ and $E_{0}(t)=E_0 \cos^2(\omega
(t-T)/8)$ for $T<t<3T$ with T the period of the field. The
threshold for over-the-barrier-ionization in neutral $N_{2}$ with an
ionization energy of $I_{p1}=0.5728$~a.u. is reached at a field
strength of $E = 0.075 \mathrm{a.u.}$ (corresponding to roughly
$2\times10^{14} \mathrm{W/cm}^2$).

Above $2\times10^{14} \mathrm{W/cm}^2$ the laser field allows an
unhindered electron escape and therefore the initial phase space is
modeled by a double electron microcanonical distribution
\cite{MengPRA89}. However, it is important to note that in
setting-up the initial phase space distribution we transition
smoothly from the tunneling to the over-the-barrier intensity
regime. Namely, we assign a random number to the phase $\phi$ of the
laser field when the first electron is ionized, see
\cite{BrabecPRA96,YePRL2008}. If the phase $\phi$ corresponds to an
instantaneous strength of the laser field $E(\phi)$ that leaves the
electron below the barrier then we use the initial conditions
dictated by the tunneling model. If the instantaneous field strength
pushes the barrier below the $I_{p1}$ of that electron then we use
the microcanonical distribution to set-up the initial phase space
distribution. This choice of initial conditions has proven
successful in past studies \cite{YePRL2008} in modeling the
experimental ratio of double versus single ionization for long laser
pulses  \cite{CornaggiaPRA2000}. With our approach we ensure a
smooth transition of the initial phase space distribution as we
change the intensity. Even at an intensity of
3$\times10^{14}\mathrm{W/cm^{2}}$ still about 75\% of the double
ionization probability corresponds to trajectories initialized using
the tunneling model, while 25\% of the probability corresponds to
trajectories initialized using the microcanonical distribution.

After setting-up the initial phase space distribution we transform
to a new system of coordinates, the so called ``regularized"
coordinates \cite{KustaanheimoJRAM65}. This transformation is exact
and explicitly eliminates the Coulomb singularity. This step is more
challenging for molecular systems since one has to ``regularize"
with respect to more than one atomic centers versus one atomic
center for atoms.  We regularize using the global regularization
scheme described in ref.~\cite{HeggieCelMech74}. Finally, we use the
Classical Trajectory Monte Carlo (CTMC) method for the time
propagation \cite{AbrinesProcPhysSoc66c}. The propagation involves
the 3-d four-body Hamiltonian in the laser field with ``frozen"
nuclei:
\begin{eqnarray*}
    H = \sum_{i=1}^{2} [\frac{p_{i}^{2}}{2}-\frac{1}{|\vec{R/2}-\vec{r}_{i}|}-\frac{1}{|-\vec{R/2}-\vec{r}_{i}|}]\\
      +\frac{1}{|\vec{r}_{1}-\vec{r}_{2}|}+(\vec{r}_{1}+\vec{r}_{2}) \cdot \vec{E}(t),
\end{eqnarray*}
where $E(t)$ is the laser electric field polarized along the
z direction and further defined as detailed above, and $\vec{R}$ is
the internuclear distance.

\section{Results}
Re-collision as a subcycle, true attosecond process permits
simulations using few cycle pulses to describe the main signatures
of electron correlation that are manifested in multi-cycle pulse
experiments \cite{StaudtePRL2007b}. Details of electron correlation
as the branching ratio of the different NSDI pathways do depend on
the pulse duration \cite{PrauznerPRA2005}. In few cycle pulses the
dominant channel of NSDI is the simultaneous ejection (SE) of both
electrons upon impact of the recolliding electron. The simultaneous
ejection channel is responsible for the NSDI hallmark: a double hump
structure in the sum of the momenta parallel to the laser field
around 4$\sqrt{U_{p}}$. Here, 2$\sqrt{U_{p}}$ is the maximum
velocity an electron can acquire from its interaction with the field
and with $U_{p}=E_{0}^{2}/(4\omega^{2})$ being the ponderomotive
energy.

\begin{figure}
  %  \centering
    \includegraphics[width=0.45\textwidth,keepaspectratio]{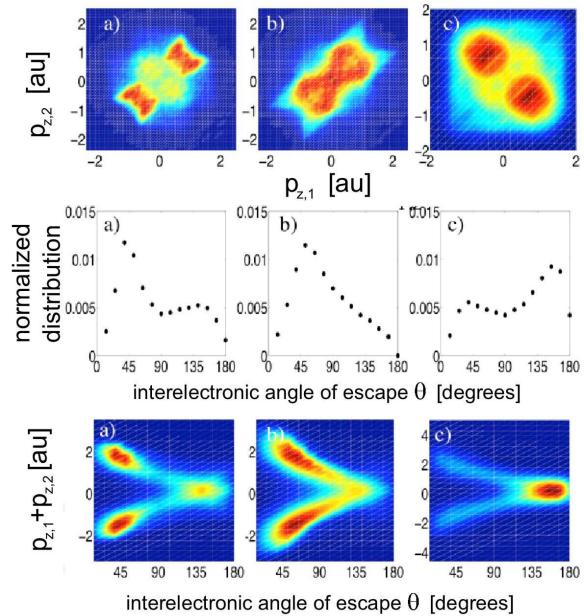}
    \caption{\label{fig2:pz1pz2all} Intensity dependence of electron correlation in the
    double ionization of N$_2$ in a 3-cycle, 800~nm pulse.
    Columns a)-c) correspond to increasing intensities 10$^{14}\mathrm{W/cm^{2}}$, 1.5$\times10^{14}\mathrm{W/cm^{2}}$,
    3$\times10^{14}\mathrm{W/cm^{2}}$, respectively.
    \textbf{Top row:} correlated momenta parallel to the polarization axis; \textbf{Middle row:}
    distribution of the inter-electronic angle of escape $\theta$;
    \textbf{Bottom row:} sum of the parallel momenta vs. $\theta$.}
\end{figure}

As the laser intensity is increased from $1\times10^{14}
\mathrm{W/cm}^2$ to $3\times10^{14} \mathrm{W/cm}^2$ a transition
takes place as evidenced by the correlated momenta of the two
escaping electrons shown in \fig{fig2:pz1pz2all}a)-c). At
$10^{14} \mathrm{W/cm}^2$ both electrons escape with very similar
momenta (\fig{fig2:pz1pz2all}a). Already at $1.5\times 10^{14}
\mathrm{W/cm}^2$ the distribution changes significantly and develops
an X-like shape (\fig{fig2:pz1pz2all}b). However, the
inter-electronic angle of escape $\theta$ (the angle between the
asymptotic electron momenta) is generally less than 90$^{\circ}$. In
contrast, at $3\times 10^{14}\mathrm{W/cm^{2}}$ the inter-electronic
angle of escape becomes larger than 90$^{\circ}$. In fact, the two
electrons tend to escape with opposite momenta
(\fig{fig2:pz1pz2all}c). This transition is also evident when one
considers the distribution of $\theta$ (\fig{fig2:pz1pz2all}c) medium
panel) with a maximum around 165$^{\circ}$. We note that our model
does not account for depletion of the initial state in
the-over-the-barrier regime. For that reason we only present the
correlated momenta for intermediate intensities up to $3\times
10^{14}\mathrm{W/cm^{2}}$ where the tunneling process still
dominates as 3:1, thereby rendering the effect of depletion small.
Finally, we note parenthetically that at $3\times10^{14}
\mathrm{W/cm}^2$ over-the-barrier ionization of the first electron
becomes possible, whereas NSDI is still the dominating mechanism in
the liberation of the second electron (compare
Fig.~\ref{fig1:CornaggiaN2}).

\subsection{Tunneling regime}
We now proceed to identify the double ionization pathways for
intensities in the tunneling regime. Two main mechanisms for
non-sequential double ionization \cite{BaierPRA2008} are well
established: simultaneous ejection (SE) upon rescattering and
re-collision-induced excitation with subsequent field ionization
(RESI).

In the SE mechanism both electrons are simultaneously ejected with
small but equal momenta, with the result that the asymptotic
momentum components along the polarization axis are mainly dictated
by the vector potential of the laser field at the time of ejection.
This agrees with \fig{fig2:pz1pz2all}a) where the maximum of
the inter-electronic angle of escape is around 30$^{\circ}$. In RESI
the rescattering electron only excites the remaining electron which
is ionized at a subsequent maximum of the laser field
\cite{LaGattutaJPB98, KopoldPRL2000,
FeuersteinPRL2001,YudinPRA2001a}.

\begin{figure}[ht]
    \includegraphics[width=0.45\textwidth,keepaspectratio]{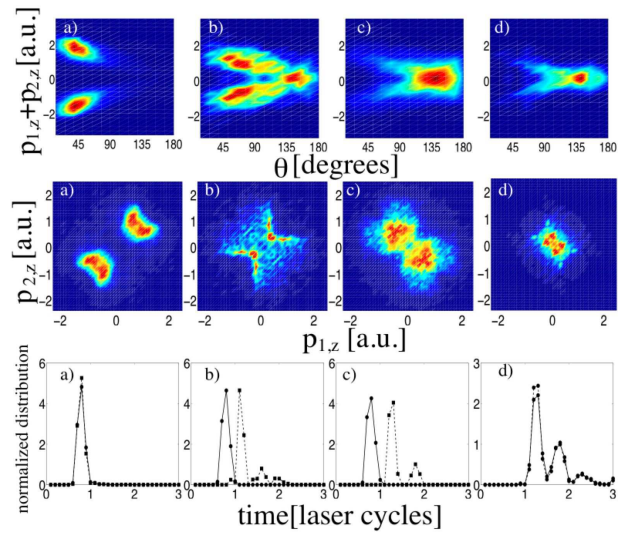}
    \caption{\label{fig3:mech} For $10^{14} \mathrm{W/cm}^{2}$ we identify
    four different double ionization mechanisms depending on each
    electron's time of ionization: the different columns refer  to a) the SE, b) the RESI , c) the NSE2 and d) the
    DE mechanisms. \textbf{Top row}: the sum of the momenta of the two
    electrons as a function of the inter-electronic angle of escape;
    \textbf{Middle row}: the correlated momenta of the two electrons; and
    \textbf{Bottom row}:  the ionization time of each electron in units of
    laser cycles.   }
\end{figure}

\begin{figure}[ht]
    \includegraphics[width=0.45\textwidth,keepaspectratio]{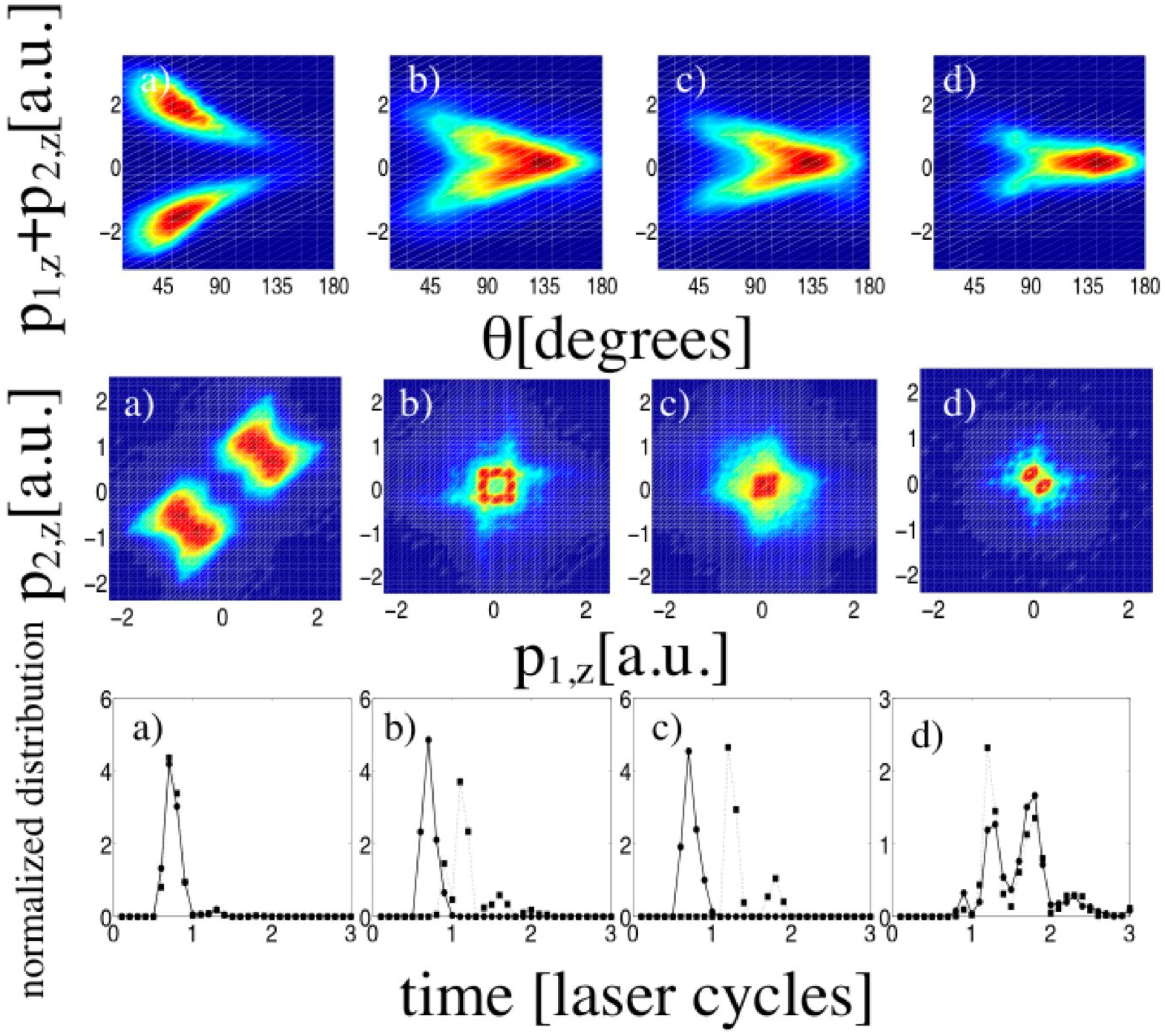}
    \caption{\label{fig4:mech} Same as for Fig.~\ref{fig3:mech} but for $1.5\times 10^{14} \mathrm{W/cm}^2$.}
\end{figure}

In Fig.~\ref{fig3:mech} and Fig.~\ref{fig4:mech} we can clearly
identify SE and RESI but also two other mechanisms. Namely, in
\fig{fig3:mech}c) we identify a NSDI pathway which starts out with
excitation upon re-collision, but in contrast to RESI the excited
electron ionizes around a subsequent zero of the field. We label
this mechanism NSE2. In \fig{fig3:mech}d) we identify yet
another mechanism that involves the formation of a doubly excited
complex at the re-collision with both electrons ionizing at a later
time as discussed in \cite{PrauznerPRA2005,BaierPRA2008}
--- we refer to this mechanism as the double excitation (DE) pathway.
Previously, it was shown how the double ionization mechanisms have
distinct traces in the spatial electron density \cite{BaierPRA2008}.
In the current work, we find that those different mechanisms also
leave distinct traces in the observable momentum space.

When electrons ionize through the SE pathway, they escape with large
momenta and with inter-electronic angles less than 90$^{\circ}$; in
ionization through the RESI pathway the sum of the parallel
components of the momenta is smaller and the electrons typically
escape with angles larger than those for the SE mechanism. When the
electrons escape through the NSE2 or the DE pathway the sum of the
momenta is zero with large inter-electronic angles of escape.
However, the two pathways have different correlated momenta as shown
in Fig.~\ref{fig3:mech} and Fig.~\ref{fig4:mech} (medium panel) ---
larger electron momenta are characteristic for the NSE2 mechanism.
The SE and NSE2 are the dominant mechanisms for a 3-cycle laser
pulse accounting for roughly 44\% and 39\%, respectively, for a
laser intensity of $10^{14} \mathrm{W/cm}^{2}$ and for 54\% and
28\%, respectively, for a laser intensity of $1.5\times10^{14}
\mathrm{W/cm}^{2}$, while RESI is around 10\% in both cases. The
contribution of the RESI and DE mechanisms increases significantly
with increasing laser pulse durations.

We note, that the identification of the different mechanisms is
based on the fact that the recolliding electron returns to the core,
i.e. on the existence of an instant of closest approach to the
remaining electron. We further verify the distinction between the
different mechanisms by analyzing the time of ionization after
re-collision shown in the bottom panels of Fig.~\ref{fig3:mech} and
Fig.~\ref{fig4:mech}. To determine the time of ionization we make use of
the ``compensated energy" \cite{LeopoldJPB79} for each electron. The
compensated energy is the electron's potential energy with respect
to the two nuclei plus its kinetic energy resulting from the
canonical momentum due to the presence of the field. Here we define
ionization as the instant when the compensated energy becomes
positive and remains positive for all subsequent times. While being
an approximate way to define the ionization time for each electron
it proves useful for the identification of the different mechanisms.

\subsection{Over-the-barrier regime}
Now we are turning towards the results obtained at an intermediate
intensity regime which is high enough that over-the-barrier
ionization of the first electron becomes possible but low enough
that the second electron cannot ionize without the energy transfer
in a re-collision with the first electron (see
\fig{fig2:pz1pz2all}c). The most striking feature of this intensity
regime is the strong anti-correlation between the electron momenta,
which resembles experimental observations by Liu \emph{et al.}
\cite{LiuPRL2008}. There, however, the anti-correlation was observed
at intensities well below the classical threshold for re-collision
double ionization. Since at these intensities a single re-collision
does not provide sufficient energy to release a second electron a
repeated energy transfer from the continuum electron via multiple
field-assisted re-collisions was held responsible for this effect.

However, in the intermediate intensity regime currently under
consideration the energy of the recolliding electron easily exceeds
the binding energy of the second electron. At $3\times 10^{14}
\mathrm{W/cm}^2$ the impact energy at re-collision can be as high as
54~eV while the ionization energy of N$_2^+$ is approximately
$I_{p2}=30\mathrm{eV}$. While this permits ionization also through
shorter trajectories, as we discuss below, experiments have shown
\cite{StaudtePRL2007b} that when excess energy is available only the
minimum energy necessary for ionization is transferred in the
collision and the recolliding electron retains most of the excess
energy.

\begin{figure}[ht]
   \includegraphics[width=0.45\textwidth,keepaspectratio]{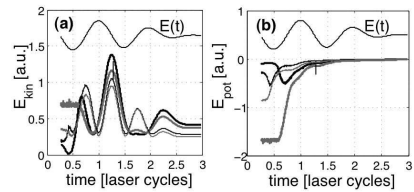}
    \caption{\label{fig5:MeanValueCollision}
    Time dependence of kinetic energy (left panel) and potential
    energy (right panel) for the recolliding electron (black lines) and the initially bound electron (grey
    lines). Shown is the average over all double ionization trajectories at $10^{14} \mathrm{W/cm}^2$ (thick lines) and
    $3\times 10^{14} \mathrm{W/cm}^2$ (thin lines).}
\end{figure}

To identify whether these ``soft" collisions are indeed responsible
for the observed anti-correlation we examine in
Fig.~\ref{fig5:MeanValueCollision} the temporal evolution of the
kinetic and potential energy of both electrons for laser intensities
$10^{14} \mathrm{W/cm}^2$ and $3\times 10^{14} \mathrm{W/cm}^2$. The
amplitude in the potential energy changes reveals that the collision
for $3\times 10^{14} \mathrm{W/cm}^2$ (thin lines) takes place
further out from the nucleus if compared to the collision at
$10^{14} \mathrm{W/cm}^2$ (thick lines). This interpretation is
further corroborated by the presence of the ``finger"-like structure
in the correlated momenta in Fig. (2)a) for $10^{14}
\mathrm{W/cm}^2$ and its absence in Fig. (2)c) for $3\times10^{14}
\mathrm{W/cm}^2$. The ``finger"-like structure was attributed in
former studies of the strongly driven He atom to strong
backscattering from the nucleus
\cite{StaudtePRL2007b,EmmanouilidouPRA2008,YePRL2008}. Finally, in
Fig.~\ref{fig5:MeanValueCollision} the temporal evolution of the
kinetic energy shows that there is a smaller relative change in the
kinetic energy of the two electrons for $3\times 10^{14}
\mathrm{W/cm}^2$ around T/2 when compared to the kinetic energy
change for  $10^{14} \mathrm{W/cm}^2$ around 3/4T.

\begin{figure}[ht]
    \includegraphics[width=0.48\textwidth,keepaspectratio]{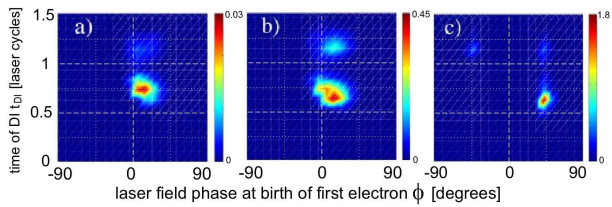}
    \caption{\label{fig6:time} Instant of double ionization $t_{DI}$ as a function
    of the phase $\phi$ of the laser field when the first electron is
    ``launched". The laser  intensity is a) $10^{14}\mathrm{W/cm^{2}}$, b)
    $1.5\times10^{14}\mathrm{W/cm^{2}}$ and c)
    $3\times10^{14}\mathrm{W/cm^{2}}$.}
\end{figure}

At $3\times 10^{14} \mathrm{W/cm}^2$  soft collisions with large
impact parameter seem to be favored due to the prevalence of short
trajectories in the total double ionization yield. In
Fig.~\ref{fig6:time} we identify the short trajectories by analyzing
the temporal structure of the re-collision process. We plot the
phase of the field when the first electron is ``launched" as a
function of the instant of double ionization (according to our
previous definition of ionization time using the compensated
energy). In the tunneling regime the recolliding electron is
``launched" shortly after the maximum of the laser field
(\fig{fig6:time}a-b), i.e. the phase $\phi$ of the laser field is
close to zero (we use a cosine field). The majority of double
ionization trajectories then recollide at around 3/4 into the laser
cycle (zero of the laser field) corresponding to long trajectories.
In contrast, in the over-the-barrier intensity regime the
recolliding electron is ``launched" when the phase of the laser
field is larger, see \fig{fig6:time}c), corresponding to short
trajectories. We find that the time delay between ``launching" of
the first electron (recolliding electron) and the first
re-collision, i.e. the maximum of the electron-electron interaction
energy, decreases from 2/3T at $1\times 10^{14} \mathrm{W/cm^2}$ to
1/2T at $3\times 10^{14} \mathrm{W/cm}^2$. Although short
trajectories have less than the maximum impact energy of $3.17U_p$ 
they can be energetic enough at high intensities to release the
second electron. However, the electron energy transfer becomes
temporally less localized when the intensity is increased to
$3\times 10^{14} \mathrm{W/cm}^2$.

\begin{figure}[ht]
    \includegraphics[width=0.3\textwidth,keepaspectratio]{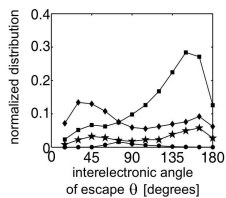}
    \caption{\label{fig7:cycles} For $3\times 10^{14} \mathrm{W/cm}^2$
    we show the distribution of the inter-electronic angle when the
    double ionization time is from T/4 to T/2 (circles); from T/2 to
    3/4T (squares); from 3/4T to T (stars); and from T to 5/4T
    (diamonds).}
\end{figure}

From the above it is clear that a transfer of energy through the
re-collision event is still necessary for the bound electron to
ionize for intensities in the intermediate regime. However, the
laser field for the higher intensities influences the motion of the
bound electron even before the re-collision process. In contrast for
the smaller intensities the asymptotic momenta components along the
polarization axis are roughly dictated by the vector potential at
the time of re-collision (around 2/3T). The significant role for
higher intensities of the laser field even before the re-collision
process can be also inferred from Fig.~\ref{fig7:cycles}. Namely,
depending on the time delay between the time of ``launching" of the
first electron and the time of double ionization, the two electrons
escape either parallel or opposite to each other. In particular, the
prevailing inter-electronic angle larger than 90$^{\circ}$ is due to
the fact that the second electron ionizes approximately T/2 after
the first electron is ``launched". With the vector potential
changing sign within half a period, this results in electrons
escaping in opposite directions.

We will now discuss the limitations of our model with respect to the
high intensity results.
Our model does not incorporate sequential double ionization,
i.e. tunneling of the second electron independent of the
``launching" of the first electron. The independent tunneling of
both electrons is known to yield uncorrelated electron momenta
\cite{WeberNature2000}, which would obscure the observed
anti-correlation. However, at $3\times 10^{14} \mathrm{W/cm}^2$
sequential double ionization can be neglected (compare
Fig.~\ref{fig1:CornaggiaN2}).

Further, our simulation does not include spatial intensity averaging
as is virtually unavoidable in the experiment. Contribution of lower
peak intensities to the overall DI signal would partially mask the
anti-correlation. This effect is expected to become increasingly
important for higher intensities.

Finally, one may legitimately ask to what extend the above finding
is a molecular effect. In a recent study of the double ionization of
the strongly driven He atom \cite{EmmanouilidouPRA2008} the
correlated momenta did not exhibit a pattern of electrons ejected in
opposite directions. However, since our explanation of the
anti-correlation does not necessitate molecular structure we would
expect in principle a similar effect in atoms. Future studies in other atomic and molecular species
can explore further whether the anti-correlation effect is generally observed for intermediate intensities
where there is an overlap of the over-the-barrier intensity regime and the regime where non-sequential ionization dominates over the sequential ionization.

\section{Conclusion}
In conclusion, we have studied the non-sequential double ionization
up to intensities where our classical model permits an intuitive
definition of re-collision. In the tunneling regime we identify two
new double ionization mechanisms and devise asymptotic observables
with distinct traces of these ionization mechanisms. For intensities
corresponding to the over-the-barrier ionization but still within
the non-sequential double ionization regime we surprisingly find
that both electrons are ejected with opposite momenta. In this intermediate intensity regime it is the field interaction with the
second, bound electron that permits soft re-collisions to assist the second
ionization step. The combination of soft re-collision and timing between the two
ionization steps forms the underlying mechanism for the observed anti-correlation of the electron momenta.
An anti-correlation pattern
was also observed in \cite{LiuPRL2008} for small intensities where the mechanism
responsible was identified as multiple inelastic field assisted recollisions. These results motivate a revision of the common notion that non-sequential
double ionization necessarily means emission of both electrons with identical
momenta.

\section{Acknowledgements}
The authors gratefully acknowledge discussions with Paul Corkum,
Andr\'{e} Bandrauk and Carla Figueira de Morrison Faria.

%\bibliographystyle{unsrt}
%\bibliography{MasterBibliography}
% \end{document}

 \end{document}